# Personalized Social Recommendations - Accurate or Private?

Ashwin Machanavajjhala
Yahoo! Research
Santa Clara, CA, USA
mvnak@yahoo-inc.com

Aleksandra Korolova[*]
Stanford University
Stanford, CA, USA
korolova@cs.stanford.edu

Atish Das Sarma[†]
Georgia Institute of Tech.
Altanta, GA, USA
dassarma@google.com

## ABSTRACT

With the recent surge of social networks such as Facebook, new forms of recommendations have become possible – recommendations that rely on one's social connections in order to make personalized recommendations of ads, content, products, and people. Since recommendations may use sensitive information, it is speculated that these recommendations are associated with privacy risks. The main contribution of this work is in formalizing trade-offs between accuracy and privacy of personalized social recommendations.

We study whether "social recommendations", or recommendations that are *solely* based on a user's social network, can be made without disclosing sensitive links in the social graph. More precisely, we quantify the loss in utility when existing recommendation algorithms are modified to satisfy a strong notion of privacy, called differential privacy. We prove lower bounds on the minimum loss in utility for any recommendation algorithm that is differentially private. We then adapt two privacy preserving algorithms from the differential privacy literature to the problem of social recommendations, and analyze their performance in comparison to our lower bounds, both analytically and experimentally. We show that good private social recommendations are feasible only for a small subset of the users in the social network or for a lenient setting of privacy parameters.

## 1. INTRODUCTION

Making recommendations or suggestions to users in order to increase their degree of engagement is a common practice for websites. For instance, YouTube recommends videos, Amazon suggests products, and Netflix recommends movies, in each case with the goal of making as *relevant* a recommendation to the user as possible. The phenomenal participation of users in social networks such as Facebook, MySpace, and LinkedIn, has given tremendous hope for designing a new type of user experience, the *social* one. The feasibility of social recommendations has been fueled by initiatives such as Facebook's Open Graph API and Google's Social Graph API, that explicitly create an underlying graph where people, events, movies, etc., are uniformly represented as nodes, and connections, such as friendship relationships, event participation, interest in a book or a movie, are represented as edges between those nodes. The connections can be established through friendship requests, event RSVPs, and social plug-ins[1], such as the "Like" button.

Recommendations based on social connections are especially crucial for engaging users who have seen very few movies, bought only a couple of products, or never clicked on ads. While traditional recommender systems default to generic recommendations, a social-network aware system can provide recommendations based on active friends. There has been much research and industrial activity to solve two problems: (a) recommending content, products, ads not only based on the individual's prior history but also based on the likes and dislikes of those the individual trusts [2, 15], and (b) recommending others whom the individual might trust [11]. In this work, we focus on recommendation algorithms based exclusively on graph link-analysis, i.e. algorithms that rely on underlying connections between people, and other entities, rather than their individual features.

Improved *social* recommendations come at a cost – they can potentially lead to a *privacy breach* by revealing sensitive information. For instance, if you only have one friend, a social recommendation algorithm that recommends to you only the products that your friends buy, would reveal the entire shopping history of that friend - information that he probably did not mean to share. Moreover, a system that uses only *trusted edges* in friend suggestions may leak information about lack of trust along specific edges, which would also constitute a privacy breach.

In this paper, we present the first theoretical study of the privacy-utility trade-offs in personalized graph link-analysis based social recommender systems. There are many different settings in which social recommendations may be used (friend, product, interest recommendations, or trust propagation), each having a slightly different formulation of the privacy concerns (the sensitive information is different in each case). However, all these problems have a common structure – recommendations are made based on a social graph (consisting of people and other entities), where some subset of edges are sensitive. For clarity of exposition, we ignore scenario specific constraints, and focus on a generic model. Our results on privacy-utility trade-offs are simple and not unexpected. The main contributions are intuitive and precise trade-off results between privacy and utility for a clear formal model of personalized social recommendations, emphasizing impossibility of social recommendation algorithms that are both accurate and private for all users.

---

[*]Supported by Cisco Systems Stanford Graduate Fellowship, Award IIS-0904325, and a gift from Cisco. Part of this work was done while interning at Yahoo! Research.

[†]Work done while at Georgia Institute of Technology and interning at Yahoo! Research.

---

[1]http://developers.facebook.com/plugins

**Our Contributions.** We consider a graph where all edges are sensitive, and an algorithm that recommends a single node $v$ to some target node $u$. We assume that the algorithm is based on a *utility function* (satisfying certain natural properties (Section 4.1)) that encodes the "goodness" of recommending each node in the graph to this target node. We focus on graph link-analysis recommenders; hence, the utility function must only be a function of the nodes and edges in the graph. Suggestions for graph link-analysis based utility functions include number of common neighbors, number of weighted paths, and PageRank distributions [12, 14]. We consider an attacker who wishes to deduce the existence of a single edge $(x, y)$ in the graph with $n$ nodes by passively observing a recommendation $(v, u)$. We measure the privacy of the algorithm using $\epsilon$-*differential privacy* - requiring the ratio of the likelihoods of the algorithm recommending $(v, u)$ on the graphs with, and without, the edge $(x, y)$, respectively, to be bounded by $e^\epsilon$. We define accuracy of a recommendation algorithm $R$ as the ratio between $R$'s expected utility to the utility achieved by an optimal (non-private) recommender. In this setting:

- We present and quantify a trade-off between accuracy and privacy of any social recommendation algorithm that is based on any general utility function. This trade-off shows a lower bound on the privacy parameter $\epsilon$ that must be incurred by an algorithm that wishes to guarantee any constant-factor approximation of the maximum possible utility. (Section 4.2)
- We present stronger lower bounds on privacy and the corresponding upper bounds on accuracy for algorithms based on two particular utility functions previously suggested for social recommendations – number of common neighbors and weighted paths [11, 12, 14]. If privacy is to be preserved when using the common neighbors utility function, only nodes with $\Omega(\log n)$ neighbors can hope to receive accurate recommendations. (Section 5)
- We adapt two well-known privacy-preserving algorithms from the differential privacy literature for the problem of social recommendations. The first (Laplace), is based on adding random noise drawn from a Laplace distribution to the utility vector [8] and then recommending the highest utility node. The second (Exponential), is based on exponential smoothing [19]. (Section 6)
- We perform experiments on two real graphs using several utility functions. The experiments compare the accuracy of Laplace and Exponential mechanisms, and the upper bound on achievable accuracy for a given level of privacy, as per our proof. Our experiments suggest three takeaways: (i) For most nodes, the lower bounds imply harsh trade-offs between privacy and accuracy when making social recommendations; (ii) The more natural Laplace algorithm performs as well as Exponential; and (iii) For a large fraction of nodes, the gap between accuracy achieved by Laplace and Exponential mechanisms and our theoretical bound is not significant. (Section 7)
- We briefly consider the setting when an algorithm may not know (or be able to compute efficiently) the entire utility vector, and propose and analyze a sampling based linear smoothing algorithm that does not require all utilities to be pre-computed (Appendix F). We conclude by mentioning several directions for future work. (Section 8)

We now discuss related work and systems, and then formalize our model and problem statement in Section 3.

## 2. RELATED WORK

Several papers propose that social connections can be effectively utilized for enhancing online applications [2, 15]. Golbeck [9] uses the trust relationships expressed through social connections for personalized movie recommendations. Mislove et al. [20] attempt an integration of web search with social networks and explore the use of trust relationships, such as social links, to thwart unwanted communication [21]. Approaches incorporating trust models into recommender systems are gaining momentum [22, 26, 27]. In practical applications, the most prominent example of graph link-based recommendations is Facebook's recommendation system that recommends to its users Pages corresponding to celebrities, interests, events, and brands, based on the social connections established in the people and Pages social graph[2]. More than 100,000 other online sites [3], including Amazon[4] and the New York Times, are utilizing Facebook's Open Graph API and social plug-ins. Some of them rely on the social graph data provided by Facebook as the sole source of data for personalization. Depending on the website's focus area, one may wish to benefit from personalized social recommendations when using the site, while keeping one's own usage patterns and connections private - a goal whose feasibility we analyze in this work.

There has been recent work discussing privacy of recommendations, but it does not consider the social graph. Calandrino et al. [5] demonstrate that algorithms that recommend products based on friends' purchases have very practical privacy concerns. McSherry and Mironov [18] show how to adapt the leading algorithms used in the Netflix prize competition to make privacy-preserving movie recommendations. Aïmeur et al. [1] propose a system for data storage for privacy-preserving recommendations. Our work differs from all of these by considering the privacy/utility trade-offs in graph-link analysis based social recommender systems, where the graph links are private.

Bhaskar et al. [4] consider mechanisms analogous to the ones we adapt, for an entirely different problem of making private frequent item-set mining practically efficient, with distinct utility notion, analysis, and results.

## 3. MODEL

This section formalizes the problem definition and initiates the discussion by describing what a social recommendation algorithm entails. We subsequently state the chosen notion of privacy, *differential privacy*. Finally, we define the accuracy of an algorithm and state the problem of designing a private and accurate social recommendation algorithm.

---

[2] http://www.facebook.com/pages/browser.php
[3] http://developers.facebook.com/blog/post/382
[4] https://www.amazon.com/gp/yourstore?ie=UTF8&ref_=pd_rhf_ys

## 3.1 Social Recommendation Algorithm

Let $G = (V, E)$ be the graph that describes the network of connections between people and entities, such as products purchased. Each recommendation is an edge $(i, r)$, where node $i$ is recommended to the *target node* $r$. Given graph $G$, and target node $r$, we denote the utility of recommending node $i$ to node $r$ by $u_i^{G,r}$, and since we are considering the graph as the sole source of data, the utility is some function of the structure of $G$. We assume that a recommendation algorithm $R$ is a probability vector on all nodes, where $p_i^{G,r}(R)$ denotes the probability of recommending node $i$ to node $r$ in graph $G$ by the specified algorithm $R$. We consider algorithms aiming to maximize the expected utility $\sum_i u_i^{G,r} \cdot p_i^{G,r}(R)$ of each recommendation. Our notation defines algorithms as probability vectors, thus capturing randomized algorithms; note that all deterministic algorithms are special cases. For instance, an obvious candidate for a recommendation algorithm would be $\mathcal{R}_{best}$ that always recommends the node with the highest utility (equivalent to assigning probability 1 to the node with the highest utility). Note that no algorithm can attain a higher expected utility of recommendations than $\mathcal{R}_{best}$.

When the graph $G$ and the target node $r$ are clear from context, we drop $G$ and $r$ from the notation – $u_i$ denotes utility of recommending $i$, and $p_i$ denotes the probability that algorithm $R$ recommends $i$. We further define $u_{\max} = \max_i u_i$, and $d_{\max}$ - the maximum degree of a node in $G$.

## 3.2 Privacy definition

Although there are many notions of privacy that have been considered in the literature, since privacy protections are extremely important in social networks, in this work we use a strong definition of privacy, *differential privacy* [7]. It is based on the following principle: an algorithm preserves privacy of an entity if the algorithm's output is not sensitive to the presence or absence of the entity's information in the input data set. In our setting of graph link-analysis based social recommendations, we wish to maintain the presence (or absence) of an edge in the graph private.

DEFINITION 1. *A recommendation algorithm $R$ satisfies $\epsilon$-differential privacy if for any pair of graphs $G$ and $G'$ that differ in one edge (i.e., $G = G' + \{e\}$ or vice versa) and every set of possible recommendations $S$,*

$$Pr[R(G) \in S] \leq exp(\epsilon) \times Pr[R(G') \in S] \quad (1)$$

*where probabilities are over random coin tosses of $R$.*

Differential privacy has been widely used in the privacy literature [3, 8, 17, 19].

In this paper we show trade-offs between utility and privacy for algorithms making a *single* social recommendation. Restricting our analysis to algorithms making one recommendation allows us to relax the privacy definition. We require Equation 1 to hold only for edges $e$ that are not incident to the node receiving the recommendation. This relaxation reflects the natural setting in which the node receiving the single recommendation (the attacker) already knows whether or not it is connected to other nodes in the graph, and hence we only need to protect the knowledge about the presence or absence of edges that don't originate from the attacker node. While we consider algorithms making a single recommendation throughout the paper, we use the relaxed variant of differential privacy *only* in Sections 5 and 7.

## 3.3 Problem Statement

We define the *private social recommendation problem* as follows. Given utility vectors (one per target node), determine a recommendation algorithm that (a) satisfies the $\epsilon$-differential privacy constraints and (b) maximizes the accuracy of recommendations. We define *accuracy* of an algorithm before formalizing our problem. For simplicity, we focus on the problem of making recommendations for a fixed target node $r$. Therefore, the algorithm takes as input only one utility vector $\vec{u}$, corresponding to utilities of recommending each of the nodes in $G$ to $r$, and returns one probability vector $\vec{p}$ (which may depend on $\vec{u}$).

DEFINITION 2 (ACCURACY). *The accuracy of an algorithm $R$ is defined as $\min_{\vec{u}} \frac{\sum u_i p_i}{u_{\max}}$.*

In other words, an algorithm is $(1 - \delta)$-accurate if (1) for every input utility vector $\vec{u}$, the output probabilities $p_i$ are such that $\frac{\sum u_i p_i}{u_{\max}} \geq (1-\delta)$, and (2) there exists an input utility vector $\vec{u}$ such that the output $p_i$ satisfies $\frac{\sum u_i p_i}{u_{\max}} = (1-\delta)$. The second condition is added for notational convenience (so that an algorithm has a well defined accuracy). In choosing the definition of accuracy, we follow the paradigm of worst-case performance analysis from the algorithms literature; average-case accuracy analysis may be an interesting direction for future work.

Recall that $u_{\max}$ is the maximum utility achieved by any algorithm (in particular by $\mathcal{R}_{best}$). Therefore, an algorithm is said to be $(1 - \delta)$-accurate if for any utility vector, the algorithm's expected utility is at least $(1-\delta)$ times the utility of the best possible algorithm. A social recommendation algorithm that aims to preserve privacy of the edges will have to deviate from $\mathcal{R}_{best}$, and accuracy is the measure of the fraction of maximum possible utility it is able to preserve despite the deviation. Notice that our definition of accuracy is invariant to rescaling utility vectors, and hence all results we present are unchanged on rescaling utilities.

We now formalize our problem definition.

DEFINITION 3 (PRIVATE SOCIAL RECOMMENDATIONS). *Design a social recommendation algorithm $R$ with maximum possible accuracy under the constraint that $R$ satisfies $\epsilon$-differential privacy.*

## 4. GENERIC PRIVACY LOWER BOUNDS

The main focus of this paper is to theoretically determine the bounds on maximum accuracy achievable by any algorithm that satisfies $\epsilon$-differential privacy. Instead of assuming a specific graph link-based recommendation algorithm, more ambitiously we aim to determine accuracy bounds for a general class of recommendation algorithms.

In order to achieve that, we first define properties that one can expect most reasonable utility functions and recommendation algorithms to satisfy. We then present a general bound that applies to all algorithms and utility functions satisfying those properties in Section 4.2 and present tighter bounds for several concrete choices of utility functions in Section 5.

## 4.1 Properties of Utility Functions and Algorithms

We present two axioms, *exchangeability* and *concentration*, that should be satisfied by a meaningful utility function in the context of recommendations on a social network. Our axioms are inspired by work of [14] and the specific utility functions they consider: number of common neighbors, sum of weighted paths, and PageRank based utility measures.

AXIOM 1 (EXCHANGEABILITY). *Let $G$ be a graph and let $h$ be an isomorphism on the nodes giving graph $G_h$, s.t. for target node $r$, $h(r) = r$. Then $\forall i : u_i^{G,r} = u_{h(i)}^{G_h,r}$.*

This axiom captures the intuition that in our setting of graph link-analysis based recommender systems, the utility of a node $i$ should not depend on the node's identity. Rather, the utility for target node $r$ only depends on the structural properties of the graph, and so, nodes isomorphic from the perspective of $r$ should have the same utility.

AXIOM 2 (CONCENTRATION). *There exists $S \subset V(G)$, such that $|S| = \beta$, and $\sum_{i \in S} u_i \geq \Omega(1) \sum_{i \in V(G)} u_i$.*

This says there are some $\beta$ nodes that together have at least a constant fraction of the total utility. This is likely to be satisfied for small enough $\beta$ in practical contexts, as in large graphs there are usually a small number of nodes that are very good recommendations for $r$ and a long tail of those that are not. Depending on the case, $\beta$ may be a constant, or may be a function growing with the number of nodes.

We now define a property of a recommendation algorithm:

DEFINITION 4 (MONOTONICITY). *An algorithm is said to be monotonic if $\forall i, j, u_i > u_j$ implies that $p_i > p_j$.*

The monotonicity property is a very natural notion for a recommendation algorithm to satisfy. It says that the algorithm recommends a higher utility node with a higher probability than a lower utility node.

In our subsequent discussions, we only consider the class of monotonic recommendation algorithms for utility functions that satisfy the exchangeability axiom as well as the concentration axiom for a reasonable choice of $\beta$. In Appendix A we briefly mention how the lower bounds can be altered to avoid this restriction.

A running example throughout the paper of a utility function that satisfies these axioms and is often successfully deployed in practical settings [11, 14] is that of the *number of common neighbors utility function*: given a target node $r$ and a graph $G$, the number of common neighbors utility function assigns a utility $u_i^{G,r} = C(i, r)$, where $C(i, r)$ is the number of common neighbors between $i$ and $r$.

## 4.2 General Lower Bound

In this section we show a lower bound on the privacy parameter $\epsilon$ for any differentially private recommendation algorithm that (a) achieves a constant accuracy and (b) is based on any utility function that satisfies the exchangeability and concentration axioms, and the monotonicity property. We only present an overview of the proof techniques. An interested reader can find the details in Appendix B.

We explain the proof technique for the lower bound using the number of common neighbors utility metric. Let $r$ be the target node for a recommendation. The nodes in any graph can be split into two groups – $V_{hi}^r$, nodes which have a high utility for the target node $r$ and $V_{lo}^r$, nodes that have a low utility. In the case of common neighbors, all nodes $i$ in the 2-hop neighborhood of $r$ (who have at least one common neighbor with $r$) can be part of $V_{hi}^r$ and the rest - of $V_{lo}^r$. Since the recommendation algorithm has to achieve a constant accuracy, it has to recommend one of the high utility nodes with constant probability.

By the concentration axiom, there are only a few nodes in $V_{hi}^r$, but there are many nodes in $V_{lo}^r$; in the case of common neighbors, node $r$ may only have 10s or 100s of 2-hop neighbors in a graph of millions of users. Hence, there exists a node $i$ in the high utility group and a node $\ell$ in the low utility group such that $\Gamma = p_i/p_\ell$ is very large ($\Omega(n)$). At this point, we show that we can carefully modify the graph $G$ by adding and/or deleting a small number ($t$) of edges in such a way that the node $\ell$ with the smallest probability of being recommended in $G$ becomes the node with the highest utility in $G'$ (and, hence, by monotonicity, the node with the highest probability of being recommended). By the exchangeability axiom, we can show that there always exist some $t$ edges that make this possible. For instance, for common neighbors utility, we can do this by adding edges between a node $i$ and $t$ of $r$'s neighbors, where $t > \max_i C(i, r)$. It now follows from differential privacy that

$$\epsilon \geq \frac{1}{t} \log \Gamma.$$

More generally, let $c$ be a real number in $(0, 1)$, and let $V_{hi}^r$ be the set of nodes $1, \ldots, k$ each of which have utility $u_i > (1-c)u_{\max}$, and let $V_{lo}^r$ be the nodes $k+1, \ldots, n$ each of which have utility $u_i \leq (1-c)u_{\max}$ of being recommended to target node $r$. Recall that $u_{\max}$ is the utility of the highest utility node. Let $t$ be the number of edge alterations (edge additions or removals) required to turn a node with the smallest probability of being recommended from the low utility group $V_{lo}^r$ into the node of maximum utility in the modified graph.

The following lemma states the main trade-off relationship between the accuracy parameter $1 - \delta$ and the privacy parameter $\epsilon$ of a recommendation algorithm:

LEMMA 1. $\epsilon \geq \frac{1}{t}\left(\ln(\frac{c-\delta}{\delta}) + \ln(\frac{n-k}{k+1})\right)$

This lemma gives us a lower bound on the privacy guarantee $\epsilon$ in terms of the accuracy parameter $1 - \delta$. Equivalently, the following corollary presents the result as an **upper bound on accuracy** that is achievable by any $\epsilon$ differential privacy preserving social recommendation algorithm:

COROLLARY 1. $1 - \delta \leq 1 - \frac{c(n-k)}{n-k+(k+1)e^{\epsilon t}}$

Consider an example of a social network with 400 million nodes, i.e., $n = 4 \cdot 10^8$. Assume that for $c = 0.99$, we have $k = 100$; this means that there are at most 100 nodes that have utility close to the highest utility possible for $r$. Recall that $t$ is the number of edges needed to be changed to make a low utility node into the highest utility node, and consider $t = 150$ (which is about the average degree in some social networks). Suppose we want to guarantee 0.1-differential privacy, then we compute the bound on the accuracy $1 - \delta$ by plugging in these values in Corollary 1. We get $(1 - \delta) \leq 1 - \frac{3.96*10^8}{4*10^8+3.33*10^8} \approx 0.46$. This suggests that for a differential privacy guarantee of 0.1, no algorithm can guarantee an accuracy better than 0.46.

Using the concentration axiom with parameter $\beta$ we prove:

LEMMA 2. *For $(1-\delta) = \Omega(1)$ and $\beta = o(n/\log n)$,*

$$\epsilon \geq \frac{\log n - o(\log n)}{t} \quad (2)$$

This expression can be intuitively interpreted as follows: in order to achieve good accuracy with a reasonable amount of privacy (where $\epsilon$ is independent of $n$), either the number of nodes with high utility needs to be very large (i.e. $\beta$ needs to be very large, $\Omega(n/\log n)$), or the number of steps needed to bring up any node's utility to the highest utility needs to be large (i.e. $t$ needs to be large, $\Omega(\log n)$).

Lemma 2 will be used in Section 5 to prove stronger lower bounds for two well studied specific utility functions, by proving tighter upper bounds on $t$, which imply tighter lower bounds for $\epsilon$. We now present a generic lower bound that applies to *any* utility function.

THEOREM 1. *For a graph with maximum degree $d_{\max} = \alpha \log n$, a differentially private algorithm can guarantee constant accuracy (approximation to utility) only if*

$$\epsilon \geq \frac{1}{\alpha}\left(\frac{1}{4} - o(1)\right) \quad (3)$$

As an example, the theorem implies that for any utility function that satisfies exchangeability and concentration (with any $\beta = o(n/\log n)$), and for a graph with maximum degree $\log n$, there is no 0.24-differentially private algorithm that achieves any constant accuracy.

**Extensions to the model.** Our results can be generalized to algorithms that do not satisfy monotonicity, algorithms providing multiple recommendations, and to settings in which we are interested in preserving node identity privacy. We refer the reader to Appendix A for details.

## 5. SPECIFIC UTILITY LOWER BOUNDS

In this section, we start from Lemma 2 and prove stronger lower bounds for particular utility functions using tighter upper bounds on $t$. Proof details are in Appendix C.

### 5.1 Privacy bound for Common Neighbors

Consider a graph and a target node $r$. We can make any node $x$ have the highest utility by adding edges from it to all of $r$'s neighbors. If $d_r$ is $r$'s degree, it suffices to add $t = d_r + O(1)$ edges to make a node the highest utility node. We state the theorem for a generalized version of common neighbors utility function.

THEOREM 2. *Let U be a utility function that depends only on and is monotonically increasing with $C(x, y)$, the number of common neighbors between $x$ and $y$. A recommendation algorithm based on U that guarantees any constant accuracy for target node $r$ has a lower bound on privacy given by $\epsilon \geq \frac{1-o(1)}{\alpha}$ where $d_r = \alpha \log n$.*

As we will show in Section 7, this is a very strong lower bound. Since a significant fraction of nodes in real-world graphs have small $d_r$ (due to a power law degree distribution), we can expect no algorithm based on common neighbors utility to be both accurate on most nodes and satisfy differential with a reasonable $\epsilon$. Moreover, this is contrary to the commonly held belief that one can eliminate privacy risk by connecting to a few high degree nodes.

Consider an example to understand the consequence of this theorem of a graph on $n$ nodes with maximum degree $\log n$. Any algorithm that makes recommendations based on the common neighbors utility function and achieves a constant accuracy is *at best*, 1.0-differentially private. Specifically, for example, such an algorithm cannot guarantee a 0.999-differential privacy on this graph.

### 5.2 Privacy bound for Weighted Paths

A natural extension of the common neighbors utility function and one whose usefulness is supported by the literature [14], is the weighted path utility function, defined as $\mathbf{score}(s, y) = \sum_{l=2}^{\inf} \gamma^{l-2} |paths^{(l)}_{(s,y)}|$, where $|paths^{(l)}_{(s,y)}|$ denotes the number of length $l$ paths from $s$ to $y$. Typically, one would consider using small values of $\gamma$, such as $\gamma = 0.005$, so that the weighted paths score is a "smoothed version" of the common neighbors score.

Again, let $r$ be the target node of degree $d_r$. We can show that the upper bound for the parameter $t$ used in Lemma 2 for a weighted paths based utility functions with parameter $\gamma$ is $t \leq (1 + o(1))d_r$, if $\gamma = o(\frac{1}{d_{\max}})$. Hence,

THEOREM 3. *A recommendation algorithm based on the weighted paths utility function with $\gamma = o(\frac{1}{d_{\max}})$ that guarantees constant accuracy for target node $r$ has a lower bound on privacy given by $\epsilon \geq \frac{1}{\alpha}(1 - o(1))$, where $d_r = \alpha \log n$.*

Notice that in Theorem 3, we get essentially the same bound as in Theorem 2 as long as the path weight parameter $\gamma$ times the maximum degree is asymptotically growing. So the same example as before suggests roughly that for nodes with at most logarithmic degree, a recommendation algorithm with constant accuracy cannot guarantee anything better than constant differential privacy.

## 6. PRIVACY-PRESERVING ALGORITHMS

There has been a wealth of literature on developing differentially private algorithms [3, 8, 19]. In this section we adapt two well known privacy tools, Laplace noise addition [8] and exponential smoothing [19], to our problem. For the purpose of this section, we will assume that given a graph and a target node, our algorithm has access to (or can efficiently compute) the utilities $u_i$ for all other nodes in the graph. Recall that our goal is to compute a vector of probabilities $p_i$ such that (a) $\sum_i u_i \cdot p_i$ is maximized, and (b) differential privacy is satisfied.

Maximum accuracy is achieved by $\mathcal{R}_{best}$, the algorithm always recommending the node with the highest utility $u_{max}$. However, it is well known that any algorithm that satisfies differential privacy must recommend every node, even the ones that have zero utility, with a non-zero probability [24].

The following two algorithms ensure differential privacy:

The Exponential mechanism creates a smooth probability distribution from the utility vector and samples from it.

DEFINITION 5. ***Exponential mechanism:*** *Given nodes with utilities $(u_1, \ldots, u_i, \ldots, u_n)$, algorithm $A_E(\epsilon)$ recommends node $i$ with probability*
$e^{\frac{\epsilon}{\Delta f} u_i} / \sum_{k=1}^{n} e^{\frac{\epsilon}{\Delta f} u_k}$, *where $\epsilon \geq 0$ is the privacy parameter, and $\Delta f$ is the sensitivity of the utility function[5].*

Unlike the Exponential mechanism, the Laplace mechanism more closely mimics the optimal mechanism $\mathcal{R}_{best}$. It first adds random noise drawn from a Laplace distribution, and like the optimal mechanism, picks the node with the maximum noise-infused utility.

DEFINITION 6. ***Laplace mechanism:*** *Given nodes with utilities $(u_1, \ldots, u_i, \ldots, u_n)$, algorithm $A_L(\epsilon)$ first computes a modified utility vector $(u'_1, \ldots, u'_n)$ as follows: $u'_i = u_i + r$*

---
[5] $\Delta f = \max_r \max_{G, G': G = G'+e} ||\vec{u}^{G,r} - \vec{u}^{G',r}||$

where $r$ is a random variable chosen from the Laplace distribution with scale[6] ($\frac{\Delta f}{\epsilon}$) independently at random for each $i$. Then, $A_L(\epsilon)$ recommends node $z$ whose noisy utility is maximal among all nodes, i.e. $z = \arg\max_i u'_i$.

THEOREM 4. *Algorithms $A_L(\epsilon)$ and $A_E(\epsilon)$ guarantee $\epsilon$ differential privacy.*

Please refer to Appendix D for the proof.

$A_L$ only satisfies monotonicity in expectation; this is sufficient for our purposes, if we perform our comparisons between mechanisms and apply the bounds to $A_L$'s expected, rather than one-time, performance.

As we will see in Section 7, in practice, $A_L$ and $A_E$ achieve very similar accuracies. The Laplace mechanism may be a bit more intuitive of the two, as instead of recommending the highest utility node it recommends the node with the highest noisy utility. It is natural to ask whether the two are isomorphic in our setting, which turns out not to be the case, as we show in Appendix E by deriving a closed form expression for the probability of each node being recommended by the Laplace mechanism as a function of its utility when $n = 2$.

Finally, both algorithms we considered so far assume the knowledge of the entire utility vector. This assumption cannot always be made in social networks for various reasons, such as prohibitively expensive storage of $n^2$ utilities for graphs of several hundred million nodes. In Appendix F, we explore a simple algorithm that assumes no knowledge of the utility vector; it only assumes that sampling from the utility vector can be done efficiently.

## 7. EXPERIMENTS

In this section we present experimental results on two real-world graphs and for two particular utility functions. We compute accuracies achieved by the Laplace and Exponential mechanisms, and compare them with the theoretical upper bound on accuracy (Corollary 1) that any $\epsilon$-differentially private algorithm can hope to achieve. Our experiments suggest three takeaways: (i) For most nodes, our bounds suggest that there is an inevitable harsh trade-off between privacy and accuracy when making social recommendations, yielding poor accuracy for most nodes under reasonable privacy parameter $\epsilon$; (ii) The more natural Laplace mechanism performs as well as the Exponential mechanism; and (iii) For a large fraction of nodes, the accuracy achieved by Laplace and Exponential mechanisms is close to the best possible accuracy suggested by our theoretical bound.

### 7.1 Experimental Setup

We use two publicly available social networks – Wikipedia vote network ($G_{WV}$) and Twitter connections network ($G_T$). While the edges in these graphs are not private, we believe that these graphs exhibit the structure and properties typical of other private social networks.

The Wikipedia vote network ($G_{WV}$) [13] is available from Stanford Network Analysis Package[7]. Some Wikipedia users are administrators, who have access to additional technical features. Users are elected to be administrators via a public vote of other users and administrators. $G_{WV}$ consists of all users participating in the elections (either casting a vote or being voted on), since inception of Wikipedia until January 2008. We convert $G_{WV}$ into an undirected network, where each node represents a user and an edge from node $i$ to node $j$ represents that user $i$ voted on user $j$ or user $j$ voted on user $i$. $G_{WV}$ consists of 7,115 nodes and 100,762 edges.

The second data set we use ($G_T$) is a sample of the Twitter connections network, obtained from [25]. $G_{WV}$ is directed, as the "follow" relationship on Twitter is not symmetrical; consists of 96,403 nodes, 489,986 edges, and has the maximum degree of 13,181.

Similar to Section 5 we use two particular utility functions: the number of common neighbors and weighted paths (with various values of $\gamma$), motivated both by literature [14] and evidence of their practical use by many companies [11], including Facebook[8] and Twitter[9]. For the directed Twitter network, we count the common neighbors and paths by following edges out of target node $r$, although other interpretations are also possible.

We select the target nodes for whom to solicit recommendations uniformly at random (10% of nodes in $G_{WV}$ and 1% of nodes in $G_T$). For each target node $r$, we compute the utility of recommending to it each of the other nodes in the network (except those $r$ is already connected to), according to the two utility functions[10]. Then, fixing a desired privacy guarantee, $\epsilon$, given the computed utility vector $\vec{u}^r$, and assuming we will make one recommendation for $r$, we compute the expected accuracy of $\epsilon$-private recommendation for $r$. For the Exponential mechanism, the expected accuracy follows from the definition of $A_E(\epsilon)$ directly; for the Laplace mechanism, we compute the accuracy by running 1,000 independent trials of $A_L(\epsilon)$, and averaging the utilities obtained in those trials. Finally, we use Corollary 1 to compute the theoretical upper bound on accuracy we derived achievable by any $\epsilon$ privacy-preserving recommendation algorithm. Note that in our experiments, we can compute exactly the value of $t$ to use in Corollary 1 for a particular $\vec{u}^r$, which turns out to be: $t = u^r_{\max} + 1 + I_{(u^r_{\max} == d_r)}$ for common neighbors and $t = \lfloor u^r_{\max} \rfloor + 2$ for weighted paths.

### 7.2 Results

**Exponential vs Laplace mechanism:** We verified in all experiments that the Laplace mechanism achieves nearly identical accuracy as the Exponential mechanism, which confirms hypothesis of Section 6 that the differences between accuracies of two mechanisms are negligible in practice.

Now we experimentally illustrate the best accuracy one can hope to achieve using an $\epsilon$ privacy-preserving recommendation algorithm given by Corollary 1. We compare this theoretical bound to the accuracy of the Exponential mechanism (which is nearly identical to that of Laplace mechanism, and the *expected accuracy* of which can be computed more efficiently). In the following Figures 1(a), 1(b), 2(a), and 2(b), we plot accuracy $(1 − \delta)$ on the $x$-axis, and the fraction of target nodes that receive recommendations of accuracy $\leq (1 − \delta)$ on the $y$-axis (similar to CDF plots).

**Common neighbors utility function:** Figures 1(a) and 1(b) show the accuracies achieved on $G_{WV}$ and $G_T$,

---

[6]In this distribution, the pdf at $y$ is $\frac{\epsilon}{2\Delta f} \exp(-|y|\epsilon/\Delta f)$

[7]http://snap.stanford.edu/data/wiki-Vote.html

[8]http://www.insidefacebook.com/2008/03/26/facebook-starts-suggesting-people-you-may-know

[9]http://techcrunch.com/2010/07/30/twitter-who-to-follow/

[10]We approximate the weighted paths utility by considering paths of length up to 3. We omit from further consideration a negligible number of the nodes that have no non-zero utility recommendations available to them.

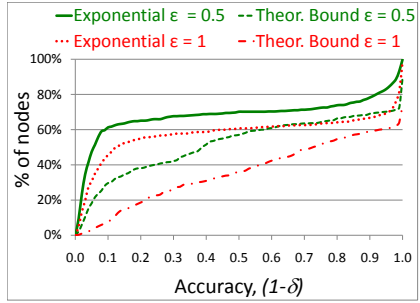
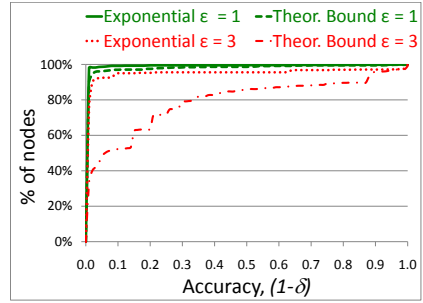

(a) On Wiki vote network  (b) On Twitter network

Figure 1: Accuracy of algorithms using # of common neighbors utility function for two privacy settings. X-axis is the accuracy $(1-\delta)$ and y-axis is the % of nodes receiving recommendations with accuracy $\leq 1-\delta$

resp., under the common neighbors utility function. As shown in Figure 1(a), for some nodes in $G_{WV}$, the Exponential mechanism performs quite well, achieving accuracy of more than 0.9. However, the number of such nodes is fairly small - for $\epsilon = 0.5$, the Exponential mechanism achieves less than 0.1 accuracy for 60% of the nodes. When $\epsilon = 1$, it achieves less than 0.6 accuracy for 60% of the nodes and less than 0.1 accuracy for 45% of the nodes. The theoretical bound proves that any privacy preserving algorithm on $G_{WV}$ will have accuracy less than 0.4 for at least 50% of the nodes, if $\epsilon = 0.5$ and for at least 30% of the nodes, if $\epsilon = 1$.

The performance worsens drastically for nodes in $G_T$ (Figure 1(b)). For $\epsilon = 1$, 98% of nodes will receive recommendations of accuracy less than 0.01, if the Exponential mechanism is used. Moreover, the poor performance is not specific to the Exponential mechanism. As can be seen from the theoretical bound, 95% of the nodes will necessarily receive less than 0.03-accurate recommendations, no matter what privacy-preserving algorithm is used. Compared to the setting of $\epsilon = 1$, the performance improves only marginally even for a much more lenient privacy setting of $\epsilon = 3$ (corresponding to one graph being $e^3 \approx 20$ times more likely than another): if the Exponential mechanism is used, more than 95% of the nodes still receive an accuracy of less than 0.1; and according to the theoretical bound, 79% of the nodes will necessarily receive less than 0.3-accurate recommendations, no matter what the algorithm.

This matches the intuition that by making the privacy requirement more lenient, one can hope to make better quality recommendations for more nodes; however, this also pinpoints the fact that for an overwhelming majority of nodes, the Exponential mechanism and any other privacy preserving mechanism can not achieve good accuracy, even under lenient privacy settings.

**Weighted paths utility function.** We show experimental results with the weighted paths utility function on $G_{WV}$ and $G_T$ in Figures 2(a) and 2(b), respectively. As expected based on discussion following proof of Theorem 3, we get a weaker theoretical bound for a higher parameter value of $\gamma$. Moreover, for higher $\gamma$, the utility function has a higher sensitivity, and hence worse accuracy is achieved by the Exponential and Laplace mechanisms.

The main takeaway is that even for a lenient $\epsilon = 1$, the theoretical and practical performances are both very poor (and worse in the case of $G_T$). For example, in $G_{WV}$, when using the Exponential mechanism (even with $\gamma = 0.0005$), more than 60% of the nodes receive accuracy less than 0.3. Similarly, in $G_T$, using the Exponential mechanism, more than 98% of nodes receive recommendations with accuracy less than 0.01. Even for a much more lenient (and, likely, unreasonable) setting of desired privacy of $\epsilon = 3$ (whose corresponding plot we omit due to space constraints), the Exponential mechanism still gives more than 98% of the nodes the same ridiculously low accuracy of less than 0.01.

Our theoretical bounds are very stringent and for a large fraction of target nodes, limits the best accuracy any privacy-preserving algorithm can hope to achieve quite severely. Even for the most lenient privacy setting of $\epsilon = 3$, at most 52% of the nodes in $G_T$ can hope for an accuracy greater than 0.5 if $\gamma = 0.05, 0.005$, or $0.0005$, and at most 24% of the nodes can hope for an accuracy greater than 0.9. These results show that even to ensure an unreasonable privacy guarantee, the utility accuracy is severely compromised.

Our findings throw into serious doubt the feasibility of developing graph link-analysis based social recommendation algorithms that are both accurate and privacy-preserving for many real-world settings.

**The least connected nodes.** Finally, in practice, it is the least connected nodes that are likely to benefit most from receiving high quality recommendations. However, our experiments suggest that the low degree nodes are also the most vulnerable to receiving low accuracy recommendations due to needs of privacy-preservation: see Figure 2(c) for an illustration of how accuracy depends on node degree.

## 8. EXTENSIONS AND FUTURE WORK

Several interesting questions remain unexplored in this work. While we have considered some particular common utility functions in this paper, it would be nice to consider others as well. Also, most works on making recommendations deal with static data. Social networks clearly change over time (and rather rapidly). This raises several issues related to changing sensitivity and privacy impacts of dynamic data. Dealing with such temporal graphs and understanding their trade-offs would be very interesting, although there is no agreement on privacy definitions for dynamic graphs.

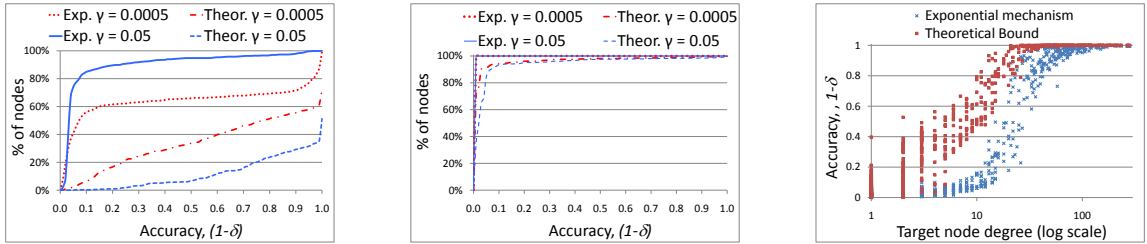

(a) Accuracy on Wiki vote network using # of weighted paths as the utility function, for $\epsilon = 1$.

(b) Accuracy on Twitter network using # of weighted paths as the utility function, for $\epsilon = 1$.

(c) Accuracy achieved by $A_E(\epsilon)$ and predicted by Theoretical Bound as a function of node degree

Figure 2: (Left, middle) Accuracy of algorithms using weighted paths utility function. X-axis is the accuracy $(1-\delta)$ and the y-axis is the % of nodes receiving recommendations with accuracy $\leq 1-\delta$ (Right) Node degree versus accuracy of recommendation (Wiki vote network, # common neighbors utility, $\epsilon = 0.5$).

Another interesting setting to consider is the case when only certain edges are sensitive. For example, in particular settings, only people-product connections may be sensitive but people-people connections are not, or users are allowed to specify which edges are sensitive. We believe our lower bound techniques could be suitably modified to consider only sensitive edges.

Finally, it would be interesting to extend our results to weaker notions of privacy than differential privacy (e.g. $k$-anonymity and relaxation of adversary's background knowledge to just the general statistics of the graph [16]).

## Acknowledgments

The authors are grateful to Arpita Ghosh and Tim Roughgarden for thought-provoking discussions; to Daniel Kifer, Ilya Mironov, and anonymous reviewers for valuable comments; and to Sergejs Melniks for help with proof of Lemma 3.

## 9. REFERENCES

[1] E. Aïmeur, G. Brassard, J. M. Fernandez, and F. S. Mani Onana. Alambic: a privacy-preserving recommender system for electronic commerce. *Int. J. Inf. Secur.*, 7(5):307–334, 2008.

[2] R. Andersen, C. Borgs, J. T. Chayes, U. Feige, A. D. Flaxman, A. Kalai, V. S. Mirrokni, and M. Tennenholtz. Trust-based recommendation systems: an axiomatic approach. In *WWW*, pages 199–208, 2008.

[3] B. Barak, K. Chaudhuri, C. Dwork, S. Kale, F. McSherry, and K. Talwar. Privacy, accuracy and consistency too: A holistic solution to contingency table release. In *PODS*, pages 273–282, 2007.

[4] R. Bhaskar, S. Laxman, A. Smith, and A. Thakurta. Discovering frequent patterns in sensitive data. In *KDD*, pages 503–512, 2010.

[5] J. Calandrino, A. Kilzer, A. Narayanan, E. Felten, and V. Shmatikov. "You might also like:" Privacy risks of collaborative filtering. In *IEEE SSP*, 2011.

[6] H. B. Dwight. *Tables of integrals and other mathematical data*. The Macmillan Company, 4th edition, 1961.

[7] C. Dwork. Differential privacy. In *ICALP*, pages 1–12, 2006.

[8] C. Dwork, F. McSherry, K. Nissim, and A. Smith. Calibrating noise to sensitivity in private data analysis. In *TCC*, pages 265–284, 2006.

[9] J. Golbeck. Generating predictive movie recommendations from trust in social networks. In *ICTM*, pages 93–104, 2006.

[10] M. Hay, C. Li, G. Miklau, and D. Jensen. Accurate estimation of the degree distribution of private networks. In *ICDM*, pages 169–178, 2009.

[11] W. Hess. People you may know, 2008. http://whitneyhess.com/blog/2008/03/30/people-you-may-know.

[12] Z. Huang, X. Li, and H. Chen. Link prediction approach to collaborative filtering. In *JCDL*, pages 141–142, 2005.

[13] J. Leskovec, D. Huttenlocher, and J.Kleinberg. Predicting positive and negative links in online social networks. In *WWW*, pages 641–650, 2010.

[14] D. Liben-Nowell and J. Kleinberg. The link prediction problem for social networks. In *CIKM*, pages 556–559, 2003.

[15] H. Ma, I. King, and M. R. Lyu. Learning to recommend with social trust ensemble. In *SIGIR*, pages 203–210, 2009.

[16] A. Machanavajjhala, J. Gehrke, and M. Goetz. Data Publishing against Realistic Adversaries. In *VLDB*, pages 790–801, 2009.

[17] A. Machanavajjhala, D. Kifer, J. Abowd, J. Gehrke, and L. Vihuber. Privacy: From theory to practice on the map. In *ICDE*, pages 277–286, 2008.

[18] F. McSherry and I. Mironov. Differentially private recommender systems: building privacy into the Netflix prize contenders. In *KDD*, pages 627–636, 2009.

[19] F. McSherry and K. Talwar. Mechanism design via differential privacy. In *FOCS*, pages 94–103, 2007.

[20] A. Mislove, K. P. Gummadi, and P. Druschel. Exploiting social networks for internet search. In *HotNets*, pages 79–84, 2006.

[21] A. Mislove, A. Post, K. P. Gummadi, and P. Druschel. Ostra: Leverging trust to thwart unwanted communication. In *NSDI*, pages 15–30, 2008.

[22] M. Montaner, B. López, and J. L. d. l. Rosa. Opinion-based filtering through trust. In *CIA*, pages 164–178, 2002.

[23] S. Nadarajah and S. Kotz. On the linear combination of laplace random variables. *Probab. Eng. Inf. Sci.*, 19(4):463–470, 2005.

[24] K. Nissim. Private data analysis via output perturbation. In *Privacy-Preserving Data Mining: Models and Algorithms*, pages 383–414. Springer, 2008.

[25] A. Silberstein, J. Terrace, B. F. Cooper, and R. Ramakrishnan. Feeding frenzy: selectively materializing users' event feeds. In *SIGMOD*, pages 831–842, 2010.

[26] G. Swamynathan, C. Wilson, B. Boe, K. Almeroth, and B. Y. Zhao. Do social networks improve e-commerce?: a study on social marketplaces. In *WOSP*, pages 1–6, 2008.

[27] C.-N. Ziegler and G. Lausen. Analyzing correlation between trust and user similarity in online communities. In *ICTM*, pages 251–265, 2004.

# APPENDIX
## A. EXTENSIONS TO THE MODEL

**Non-monotone algorithms.** Our results can be generalized to algorithms that do not satisfy the monotonicity property, assuming that they only use the utilities of nodes (and node names do not matter). We omit the exact lemmas analogous to Lemmas 1 and 2 but remark that the statements and our qualitative conclusions will remain essentially unchanged, with the exception of the meaning of variable $t$. Currently, we have $t$ as the number of edge additions or removals necessary to *make* the node with the smallest probability of being recommended into the node with the highest utility. We then argue about the probability with which the highest utility node is recommended by using monotonicity. Without the monotonicity property, $t$ would correspond to the number of edge alterations necessary to *exchange* the node with the smallest probability of being recommended and the node with the highest utility. We can then use just the exchangeability axiom to argue about the probability of recommendation. Notice that this requires a slightly higher value of $t$, and consequently results in a slightly weaker lower bound.

**Multiple recommendations.** We show that even when trying to make a single social recommendation, the results are mostly negative - i.e. there is a fundamental limit on the accuracy of privacy-preserving recommendations. Our results would imply stronger negative results for making multiple recommendations.

**Node identity privacy.** Our results can be generalized to preserving the privacy of node identities as well. Differential privacy in that case would be concerned with the ratio of the likelihoods of any recommendation on two graphs that differ in only the neighborhood of exactly one node in the graph. Unlike in the edge privacy case, where we are allowed to modify only one edge, in the node privacy case we can completely modify one node (i.e. rewire all the edges incident on it) [10]. It can be easily seen that in our lower bound proof, one can *exchange* a least useful node $v_{min}$ into the most useful node $v_{max}$ in $t = 2$ such steps – rewire the edges of $v_{min}$ to look like $v_{max}$ and vice versa. Thus for node identity privacy, we need $\epsilon \geq \frac{\log n - o(\log n)}{2}$ for constant accuracy.

## B. PROOFS FOR GENERAL BOUND

CLAIM 1. *Suppose the algorithm achieves accuracy of $(1-\delta)$ on a graph $G$. Then there exists a node $x$ in $V_{lo}^r(G)$, such that its probability being recommended is at most $\frac{\delta}{c(n-k)}$, e.g. $p_x^G \leq \frac{\delta}{c(n-k)}$.*

PROOF. In order to achieve $(1-\delta)$ accuracy, at least $\frac{c-\delta}{c}$ of the probability weight has to go to nodes in the high utility group. Denote by $p^+$ and $p^-$ the total probability that goes to high and low utility nodes, respectively, and observe that $p^+ u_{\max} + (1-c)u_{\max}p^- \geq \sum_i u_i p_i \geq (1-\delta)u_{\max}$ and $p^+ + p^- \leq 1$, hence, $p^+ > \frac{c-\delta}{c}, p^- \leq \frac{\delta}{c}$. □

**Proof of Lemma 1**

PROOF. Using the preceding Claim, let $x$ be the node in $G_1$ that is recommended with utility of at most $\frac{\delta}{c(n-k)}$ by the privacy-preserving $(1-\delta)$-accurate algorithm. And let $G_2$ be the graph obtained by addition of $t$ edges to $G_1$ chosen so as to turn $x$ into the node of highest utility. By differential privacy, we have $\frac{p_x^{G_2}}{p_x^{G_1}} \leq e^{\epsilon t}$.

In order to achieve $(1-\delta)$ accuracy on $G_2$, at least $\frac{c-\delta}{c}$ of the probability weight has to go to nodes in the high utility group, and hence by monotonicity, $p_x^{G_2} > \frac{c-\delta}{c(k+1)}$. Combining the previous three inequalities, we obtain:

$\frac{(c-\delta)(n-k)}{(k+1)\delta} = \frac{\frac{c-\delta}{c(k+1)}}{\frac{\delta}{c(n-k)}} < \frac{p_x^{G_2}}{p_x^{G_1}} \leq e^{\epsilon t}$,

hence $\epsilon \geq \frac{1}{t}\left(\ln\left(\frac{c-\delta}{\delta}\right) + \ln\left(\frac{n-k}{k+1}\right)\right)$, as desired. □

**Proof of Lemma 2**

CLAIM 2. *If $c = \left(1 - \frac{1}{\log n}\right)$, then $k = O(\beta \log n)$ where $\beta$ is the parameter of the concentration axiom.*

PROOF. Now consider the case when $c = \left(1 - \frac{1}{\log n}\right)$.

Therefore, $k$ is the number of nodes that have utility at least $\frac{u_{\max}}{\log n}$. Let the total utility mass be $U = \sum_i u_i$. Since by concentration, the $\beta$ highest utility nodes add up to a total utility mass of $\Omega(1) * U$, we have $u_{\max} \geq \Omega(\frac{U}{\beta})$. Therefore, $k$, the number of nodes with utility at least $\frac{u_{\max}}{\log n}$ is at most $\frac{U \log n}{u_{\max}}$ which is at most $O(\beta \log n)$. □

PROOF. We now prove the Lemma using Lemma 1 and Claim 2. Substituting these in the expression, if we need $1 - \frac{c(n-k)}{n-k+(k+1)e^{\epsilon t}}$ to be $\Omega(1)$, then require $(k+1)e^{\epsilon t}$ to be $\Omega(n-k)$. (Notice that if $(k+1)e^{\epsilon t} = o(n-k)$, then $\frac{c(n-k)}{n-k+(k+1)e^{\epsilon t}} \geq c - o(1)$, which is $1 - o(1)$.).

Therefore, if we want an algorithm to obtain constant approximation in utility, i.e. $(1-\delta) = \Omega(1)$, then we need the following (assuming $\beta$ to be small): $(O(\beta \log n))e^{\epsilon t} = \Omega((n - O(\beta \log n)))$, or (for small enough $\beta$), $e^{\epsilon t} = \Omega(\frac{n}{\beta \log n})$. Simplifying,

$\epsilon \geq \frac{\log n - \log \beta - \log \log n}{t}$, hence $\epsilon \geq \frac{\log n - o(\log n)}{t}$ □

**Proof of Theorem 1 (Any utility function)**

PROOF. Recall that $d_{\max}$ denotes the maximum degree in the graph. Using the exchangeability axiom, we can show that $t \leq 4d_{\max}$ in any graph. Consider the highest utility node and the lowest utility node, say $x$ and $y$ respectively. These nodes can be *interchanged* by deleting all of $x$'s current edges, adding edges from $x$ to $y$'s neighbors, and doing the same for $y$. This requires at most $4d_{\max}$ changes. By applying the upper bound on $t$ in Lemma 2 we obtain the desired result. □

## C. PROOFS FOR COMMON NEIGHBORS AND WEIGHTED PATH UTILITY

**Proof of Theorem 2 (Common Neighbors)**

PROOF. It is sufficient to prove the following upper bound on $t$.

CLAIM 3. *For common neighbors based utility functions, when recommendations for $r$ are being made, we have $t \leq d_r + 2$, where $d_r$ is the degree of node $r$.*

PROOF. Observe that if the utility function for recommendation is # of common neighbors, then one can make any zero utility node, say $x$, for source node $r$ into a max utility node by adding $d_r$ edges to all of $r$'s neighbors and additionally adding two more edges (one each from $r$ and $x$) to some node with small utility. This is because the highest utility node has at most $d_r$ common neighbors with $r$ (one of which could potentially be $x$). Further, adding these edges cannot increase the number of common neighbors for any other node beyond $d_r$. □

**Proof of Theorem 3 (Weighted Paths)**

PROOF. The number of paths of length $l$ between two nodes is at most $d_{\max}^{l-1}$. Let $x$ be the highest utility node (with utility $u_x$) and let $y$ be the node we wish to make the highest utility node after adding certain edges. If we are making recommendations for node $r$, then the maximum number of common neighbors with $r$ is at most $d_r$.

We know that $u_x \leq d_r + \sum_{l=3}^{\inf} \gamma^{l-2} d_{\max}^{l-1}$. In fact, one can tighten the second term as well.

We rewire the graph as follows. Any $(c-1)d_r$ nodes (other than $y$ and the source node $r$) are picked; here $c > 1$ is to be determined later. Both $r$ and $y$ are connected to these $(c-1)d_r$ nodes. Additionally, $y$ is connected to all of $r$'s $d_r$ neighbors. Therefore, we now get the following: $u_y \geq cd_r$

Now we wish to bound by above the utility of any other node in the network in this rewired graph. Notice that every other node still has at most $d_r$ paths of length 2 with the source. Further, there are only two nodes in the graph that have degree more than $d_{\max} + 1$, and they have degree at most $(c+1)d_{\max}$. Therefore, the number of paths of length $l$ for $l \geq 3$ for any node is at most $((c+1)d_{\max})^2 \cdot (d_{\max}+1)^{l-3}$. This can be further tightened to $((c+1)d_{\max})^2 \cdot (d_{\max})^{l-3}$. We thus get the following for any $x$ in the rewired graph: $u_x \leq d_r + (c+1)^2 \sum_{l=3}^{\infty} \gamma^{l-2} d_{\max}^{l-1}$.
Now consider the case where $\gamma < \frac{1}{d_{\max}}$. We get

$u_x \leq d_r + \frac{(c+1)^2 \gamma d_{\max}^2}{1-\gamma d_{\max}}$. We now want $u_y \geq u_x$. This reduces to $(c-1) \geq \frac{(c+1)^2 \gamma d_{\max}}{1-\gamma d_{\max}}$.

Now if $\gamma = o(\frac{1}{d_{\max}})$ then it is sufficient to have $(c-1) = \Omega(\gamma d_{\max})$ which can be achieved even with $c = 1 + o(1)$. Now notice that we only added $d_r + 2(c-1)d_r$ edges to the graph. This completes the proof of the theorem. □

**Discussion of a relationship between the common neighbors and weighted paths utility functions.**
Since common neighbors is an extreme case of weighted paths (as $\gamma \to 0$), we are able to obtain the same lower bound (up to $o(1)$ terms) when $\gamma$ is small, i.e., $\gamma \approx o(\frac{1}{d_{\max}})$. Can one obtain (perhaps weaker) lower bounds when, say, $\gamma = \Theta(\frac{1}{d_{\max}})$? Notice that the proof only needs $(c-1) \geq \frac{(c+1)^2 \gamma d_{\max}}{1-\gamma d_{\max}}$. We then get a lower bound of $\epsilon \geq \frac{1}{\alpha}(\frac{1-o(1)}{2c-1})$, where $d_r = \alpha \log n$. Setting $\gamma d_{\max} = s$, for some constant $s$, we can find the smallest $c$ that satisfies $(c-1) \geq \frac{(c+1)^2 s}{1-s}$. Notice that this gives a nontrivial lower bound (i.e. a lower bound tighter than the generic one presented in the previous section), as long as $s$ is a sufficiently small constant.

## D. PRIVACY OF LAPLACE AND EXPONENTIAL MECHANISMS

**Proof of Theorem 4**

PROOF. The proof that $A_E(\epsilon)$ guarantees $\epsilon$ differential privacy follows from McSherry et al [19].

The proof that $A_L(\epsilon)$ guarantees $\epsilon$ differential privacy follows from the privacy of Laplace mechanism when publishing histograms [8]; each node can be treated as a histogram bin and $u'_i$ is the noisy count for the value in that bin. Since $A_L(\epsilon)$ is effectively doing post-processing by releasing only the name of the bin with the highest noisy count, the algorithm remains private. □

## E. COMPARISON OF LAPLACE AND EXPONENTIAL MECHANISMS

Although we have observed in Section 7 that the Exponential and Laplace mechanisms perform comparably and know anecdotally that the two are used interchangeably in practice, the two mechanisms are not equivalent.

We compute the probability of each node being recommended by each of the mechanisms when $n = 2$, using the help of the following Lemma:

LEMMA 3. *Let $u_1$ and $u_2$ be two non-negative real numbers and let $X_1$ and $X_2$ be two random variables drawn independently from the Laplace distribution with scale $b = \frac{1}{\epsilon}$ and location 0. Assume wlog that $u_1 \geq u_2$. Then*

$$Pr[u_1 + X_1 > u_2 + X_2] = 1 - \frac{1}{2}e^{-\epsilon(u_1-u_2)} - \frac{\epsilon(u_1-u_2)}{4e^{\epsilon(u_1-u_2)}}$$

To the best of our knowledge, this is the first explicit closed form expression for this probability (the work of [23] gives a formula that does not apply to our setting).

PROOF. Let $\phi_X(u)$ denote the characteristic function of the Laplace distribution, it is known that $\phi_X(u) = \frac{1}{1+b^2u^2}$. Moreover, it is known that if $X_1$ and $X_2$ are independently distributed random variables, then

$$\phi_{X_1+X_2}(u) = \phi_{X_1}(u)\phi_{X_2}(u) = \frac{1}{(1+b^2u^2)^2}$$

Using the inversion formula, we can compute the pdf of $X = X_1 + X_2$ as follows:

$$f_X(x) = F'_X(x) = \frac{1}{2\pi}\int_{-\infty}^{\infty} e^{-iux}\phi_X(u)du$$

For $x > 0$, the pdf of $X_1 + X_2$ is $f_X(x) = \frac{1}{4b}(1+\frac{x}{b})e^{-\frac{x}{b}}$ (adapting formula 859.011 of [6]) and the cdf is $F_X(x) = 1 - \frac{1}{4}\epsilon e^{-\epsilon x}(\frac{2}{\epsilon} + x)$.

Hence $Pr[u_1 + X_1 > u_2 + X_2] = Pr[X_2 - X_1 < u_1 - u_2] = 1 - \frac{1}{4}\epsilon e^{-\epsilon(u_1-u_2)}(\frac{2}{\epsilon} + (u_1-u_2)) = 1 - \frac{1}{2}e^{-\epsilon(u_1-u_2)} - \frac{\epsilon(u_1-u_2)}{4e^{\epsilon(u_1-u_2)}}$ □

It follows from Lemma 3 and the definition of the mechanisms in Section 6 that when $n = 2$, and the node utilities are $u_1$ and $u_2$ (assuming $u_1 \geq u_2$ wlog), the Laplace mechanism will recommend node 1 with probability $1 - \frac{1}{2}e^{-\epsilon(u_1-u_2)} - \frac{\epsilon(u_1-u_2)}{4e^{\epsilon(u_1-u_2)}}$, and the exponential mechanism will recommend node 1 with probability $\frac{e^{\epsilon u_1}}{e^{\epsilon u_1}+e^{\epsilon u_2}}$. The reader can verify that the two are not equivalent through value substitution.

## F. SAMPLING AND LINEAR SMOOTHING FOR UNKNOWN UTILITY VECTORS

Both the differentially private algorithms we considered in Section 6 assume the knowledge of the entire utility vector, an assumption that cannot always be made in social networks for various reasons. Firstly, computing, as well as storing the utility of $n^2$ pairs may be prohibitively expensive when dealing with graphs of several hundred million nodes. Secondly, even if one could compute and store them, these graphs change at staggering rates, and therefore, utility vectors are also constantly changing.

We now propose a simple algorithm that assumes no knowledge of the utility vector; it only assumes that sampling from the utility vector can be done efficiently. We show how to modify any given efficient recommendation algorithm $A$, which is $\mu$-accurate but not provably private, into an algorithm $A_S(x)$ that guarantees differential privacy, while still preserving, to some extent, the accuracy of $A$.

DEFINITION 7. *Given an algorithm $A = (p_1, p_2, \ldots, p_n)$, which is $\mu$-accurate, algorithm $A_S(x)$ recommends node $i$ with probability $\frac{1-x}{n} + xp_i$, where $0 \leq x \leq 1$ is a parameter.*

Intuitively, $A_S(x)$ corresponds to flipping a biased coin, and, depending on the outcome, either sampling a recommendation using $A$ or making one uniformly at random.

THEOREM 5. *$A_S(x)$ guarantees $\ln(1 + \frac{nx}{1-x})$-differential privacy and $x\mu$ accuracy.*

PROOF. Let $p_i'' = \frac{1-x}{n} + xp_i$. First, observe that $\sum_{i=1}^n p_i'' = 1$, and $p_i'' \geq 0$, hence $A_S(x)$ is a valid algorithm. The utility of $A_S(x)$ is $U(A_S(x)) = \sum_{k=1}^n u_k p_k'' = \sum_{k=1}^n (\frac{1-x}{n})u_k + \sum_{k=1}^n xp_k u_k \geq x\mu u_{\max}$, where we use the facts that $\sum_k u_k \geq 0$ and $\sum p_k u_k \geq \mu u_{\max}$ by assumption on $A$'s accuracy. Hence, $U(A_S(x))$ has accuracy $\geq x\mu$.

For the privacy guarantee, note that $\frac{1-x}{n} \leq p_i'' \leq \frac{1-x}{n} + x$, since $0 \leq p_i \leq 1$. These upper and lower bounds on $p_i''$ hold for *any* graph and utility function. Therefore, the change in the probability of recommending $i$ for any two graphs $G$ and $G'$ that differ in exactly one edge is at most:

$$\frac{p_i(G)}{p_i(G')} \leq \frac{x + \frac{1-x}{n}}{\frac{1-x}{n}} = 1 + \frac{nx}{1-x}.$$

Therefore, $A_S$ is $\ln(1 + \frac{nx}{1-x})$-differentially private, as desired. □

Note that to guarantee $2\epsilon$-differentially privacy for $A_S(x)$, we need to set the parameter $x$ so that $\ln(1 + \frac{nx}{1-x}) = 2c \ln n$ (rewriting $\epsilon = c \ln n$), namely $x = \frac{n^{2c}-1}{n^{2c}-1+n}$.